\documentclass[conference]{IEEEtran}
\IEEEoverridecommandlockouts
\usepackage{multirow}

\IEEEoverridecommandlockouts
\usepackage{cite}
\usepackage{amsmath,amssymb,amsfonts}
\usepackage{algorithmic}
\usepackage{graphicx}
\usepackage{textcomp}
\usepackage{xcolor}
\usepackage{array}
\usepackage{listings}
\usepackage{color}
\newlength\MAX  
\newcommand*\Chart[1]{\rlap{\textcolor{black!20}{\rule{\MAX}{2ex}}}\rule{#1\MAX}{2ex}}

\def\BibTeX{{\rm B\kern-.05em{\sc i\kern-.025em b}\kern-.08em
    T\kern-.1667em\lower.7ex\hbox{E}\kern-.125emX}}
\begin{document}

\definecolor{dkgreen}{rgb}{0,0.6,0}
\definecolor{gray}{rgb}{0.5,0.5,0.5}
\definecolor{mauve}{rgb}{0.58,0,0.82}

\lstset{frame=tb,
  language=Java,
  aboveskip=1mm,
  belowskip=1mm,
  showstringspaces=false,
  columns=flexible,
  basicstyle={\small\ttfamily},
  numbers=none,
  numberstyle=\tiny\color{gray},
  keywordstyle=\color{blue},
  commentstyle=\color{dkgreen},
  stringstyle=\color{mauve},
  breaklines=true,
  breakatwhitespace=true,
  tabsize=1
}

\newcommand{\carlos}[1]{{\textcolor[rgb]{1,0,1}{Carlos: #1}}}
\newcommand{\marcelo}[1]{{\textcolor{cyan}{Marcelo escreveu: #1}}}



\title{How do Developers Improve Code Readability? An Empirical Study of Pull Requests}

\author{\IEEEauthorblockN{1\textsuperscript{st} Carlos Eduardo C. Dantas}
\IEEEauthorblockA{\textit{Federal University of Uberlândia} \\
Uberlândia, Brazil \\
carloseduardodantas@iftm.edu.br}
\and
\IEEEauthorblockN{2\textsuperscript{nd} Adriano M. Rocha}
\IEEEauthorblockA{\textit{Federal University of Uberlândia} \\
Uberlândia, Brazil \\
adriano.rocha@ufu.br}
\and
\IEEEauthorblockN{3\textsuperscript{rd} Marcelo A. Maia}
\IEEEauthorblockA{\textit{Federal University of Uberlândia} \\
Uberlândia, Brazil \\
marcelo.maia@ufu.br}
}

\maketitle

\begin{abstract}
Readability models and tools have been proposed to measure the effort to read code. However, these models are not completely able to capture the quality improvements in code as perceived by developers. To investigate possible features for new readability models and production-ready tools, we aim to better understand the types of readability improvements performed by developers when actually improving code readability, and identify discrepancies between suggestions of automatic static tools and the actual improvements performed by developers. We collected 370 code readability improvements from 284 \textit{Merged Pull Requests} (PRs) under 109 GitHub repositories and produce a catalog with 26 different types of code readability improvements, where in most of the scenarios, the developers improved the code readability to be more intuitive, modular, and less verbose. Surprisingly, SonarQube only detected 26 out of the 370 code readability improvements. This suggests that some of the catalog produced has not yet been addressed by SonarQube rules, highlighting the potential for improvement in Automatic static analysis tools (ASAT) code readability rules as they are perceived by developers. 

\end{abstract}


\maketitle

\section{Introduction}

Readability is a crucial characteristic in software development because developers often spend a significant amount of time reading and understanding code \cite{Minelli2015}, particularly when working on existing software written by other developers \cite{Erlikh2000}. According to a previous study, a survey was conducted among developers, which revealed that 83.8\% of them consider code readability to be an essential factor in their source code writing activities \cite{Piantadosi2020}. When code lacks readability, it can be difficult for other developers to understand it, and consequently fix, evolve or change it, and refactoring operations are one way to make it more comprehensible  \cite{Pantiuchina2020} \cite{Johnson2019}.  A more in-depth study on how developers improve code readability could be beneficial in understanding refactoring operations aimed at automatically improving code readability \cite{Piantadosi2022}.


There are several practices and tools that have been proposed to assess code readability. Automatic static analysis tools (ASAT), such as \textit{SonarQube}\footnote{https://www.sonarsource.com/products/sonarqube/}, offer a set of metrics that can identify potential readability violations in code. The readability models aim to measure the effort required to read code on single snapshots \cite{Scalabrino2018} \cite{Posnett2011} \cite{Buse2010}, or in code changes \cite{Roy2020}. However, assessing code readability can be challenging due to the complex syntax and semantics involved in source code \cite{Busjahn2015}. For instance, recent research  proposed a model that incorrectly classified the code readability improvement in 98 out of 297 samples (33\%) \cite{Roy2020}. Another study did not find a correlation between code understandability and 121 metrics related to code itself \cite{Scalabrino2017}. Furthermore, most readability models are unable to capture real-world improvements in code readability \cite{Pantiuchina2018}, leading to a discrepancy between models and the types of improvements made by developers \cite{Fakhoury2019}. 




Despite the challenges involved, developers continue to use tools and models to assess and improve the readability of their code. For instance, readability models are often utilized to rank readable code snippets \cite{Hora2021APISonarMA} \cite{HORA2021110971} \cite{Moreno2015}.  Additionally, developers consider the use of ASAT relevant for improving software quality \cite{Marcilio2019}. Recent research has explored the development of new readability models that can identify improvements in different versions of the same source code \cite{Roy2020}, as well as efforts to improve the accuracy of readability models \cite{Mi2018}. However, there is still lack of production-ready tools and new readability models  that effectively categorize changes in code readability,  motivating the need for further research on how developers improve the readability of their code in real-world projects \cite{Piantadosi2022}. 

In this study, we investigate the code readability improvements made by developers on  \textit{Git Pull Requests} (PRs), and compare them with the improvement suggestions identified by SonarQube ASAT. Our aim is to identify any discrepancies between suggestions of an ASAT and the actual improvements performed by developers, i.e., to assess how the improvements made by developers in PRs align with the  established code readability rules in SonarQube ASAT. This analysis will enable us to identify any potential gaps between the code readability improvements performed by developers and the established code readability rules defined in SonarQube ASAT. We decided to use PRs for identifying code readability improvements based on two main benefits:


\begin{itemize}
    \item Pull requests (PRs) promote collaborative code reviewing, where developers submit code changes and reviewers can suggest modifications before merging the code into the repository. This approach helps ensure that commits with readability improvements undergo a peer code review process \cite{Caitlin2018}.
     \item Developers often provide detailed descriptions of their code changes in PR comments to help reviewers better understand their work. These comments can offer valuable insights into the developer's own perception of readability improvements \cite{Bacchelli2013}. 
\end{itemize}

The study is driven by three main research questions:

\begin{itemize}
    \item \textbf{RQ \#1)} What types of code readability improvements do developers describe and perform in Pull Requests (PRs)?

    \item \textbf{RQ \#2)} Do developers fix code readability issues identified by SonarQube in readability PRs?
    
     \item \textbf{RQ \#3)} Can SonarQube detect the code readability improvements performed by developers in PRs?

\end{itemize} 

\textbf{Contribution:} To the best of our knowledge, this is the first study investigating code readability improvements using the developer's and reviewers' explanations on PRs. We extracted 370 code readability improvements performed by developers from 284 PRs under 109 GitHub Java-based engineering projects, producing a catalog of 26 types of code readability improvements. We also extracted the readability issues identified by SonarQube for the 370 instances, with the goal of comparing the improvements performed by developers to the recommendations provided by ASATs.

The paper is organized as follows. Section 2 discusses concepts about code readability, GitHub Pull Requests (PRs) and ASAT. Section 3 presents the methodology to answer the research questions. The results are reported and discussed in Section 4. Section 5 presents the implications for developers and researches, including the threats that could affect the validity of this study. Section 6 presents the related literature. Finally, Section 7 summarizes our observations in lessons learned and outlines directions for future work.

\section{Background}

\subsection{Code Readability}

According to Buse \& Weimer, code readability refers to how easily a human can understand code \cite{Buse2010}. Reading code is often the most time-consuming task in software maintenance \cite{Rugaber2000TheUO}, making code readability an important aspect of software development. Code readability should be assessed during software inspections, often carried out using an Modern Code Review (MCR) process \cite{Knight1991}. The definition of code readability is closely related to similar concepts such as legibility, understandability, comprehensibility, and simplicity.

Delano and colleagues define legibility as the ease of identifying program elements, and readability as a set of factors that make a program easier or harder to read \cite{Oliveira2020}. Posnett and colleagues compare readability to syntactic analysis and understandability to semantic analysis \cite{Posnett2011}, with semantic aspects including statements, beacons, and motifs \cite{Scalabrino2017}. Difficult-to-read code is also difficult to understand \cite{Boehm1976}. Rambally defines comprehensibility as the ease of maintaining, testing, and modifying code, while readability is how easy the code is to read and understand \cite{Rambally1986TheIO}. Börstler considers readability a basic prerequisite for understandability, where syntactical elements are easy to spot and recognize, and simplicity is a possible characteristic of readability \cite{Borstler2015}.


Code readability, as defined in previous works, is a crucial factor in understanding code. This encompasses syntactic aspects of the code, which influences a developer's judgment of how easy the code is to comprehend.

\subsection{Modern Code Review (MCR) and Pull Requests (PRs)}

Modern Code Review (MCR) is a lightweight and tool-assisted approach to code review. MCR is asynchronous and centered around code changes submitted by authors (i.e., developers), which are then manually examined and revised by one or more reviewers (i.e., other developers) \cite{Bacchelli2013}. These code changes can encompass a variety of improvements, such as bug fixes, new features, or refactorings \cite{Palomba2017}.

Git Pull Requests (PRs) are a fundamental part of MCR because they promote a well-defined and collaborative review process. GitHub, which hosts around 76 million developers \cite{Octoverse}, commonly uses PRs as a way to collaborate in a repository \cite{PullRequest}. To submit a PR, the author (developer) first forks a Git branch from the repository, implements and commits changes in the forked branch, and then opens a PR to submit his commit(s) for review using the MCR process. The author includes a title and description (body) of their change, and each commit has its own message. During the review process, reviewers can submit comments and request changes, and the author can respond to comments and perform new commits with requested changes. After the review process, the reviewer may approve the PR and merge the changes into a selected branch in the repository, or close the PR without merging the changes.

\subsection{Automatic static analysis tools (ASAT)}

Automatic Static analysis tools (ASAT) are designed to analyze source code without the need for running the program \cite{Nielson2010}. ASATs help to reveal coding rule violations as warnings during the development process, allowing developers to correct them before they are released as part of the software, thus ensuring a higher quality of software during the development process \cite{Johnson2013}. ASATs are particularly effective at identifying certain types of defects that may not be detected by unit tests or manual inspection \cite{Hovemeyer2004}.

ASATs can be integrated into continuous integration (CI) workflows to ensure that code quality is maintained throughout the software development process. One of the most widely adopted tools for code analysis in CI environments is SonarQube \cite{Vassallo2018}, which supports 27 programming languages in its latest version (9.2) and is used by over 200,000 companies. SonarQube comes with its own set of rules and configurations, but additional rules can also be added. Moreover, the tool incorporates rules from other static and dynamic code analysis tools, such as FindBugs and PMD\footnote{https://pmd.github.io/latest/index.html}.

 \textit{SonarQube} provides a comprehensive set of more than 600 rules for Java static code analysis\footnote{https://rules.sonarsource.com/java}, which are considered coding standards. When a piece of code violates one of these rules, an issue is raised and classified as either a bug, vulnerability, or code smell. Bugs refer to issues related to code that are demonstrably wrong. A Vulnerability occurs when a piece of code has the potential to be exploited, resulting in harm to the software. Code smells refer to instances of code that are poorly designed or structured, making it confusing and difficult to understand and maintain.

\begin{figure}[]
\centerline{\includegraphics[width=0.50\textwidth]{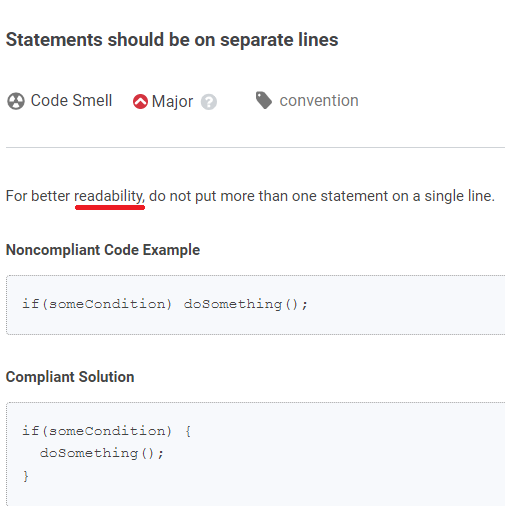}}
\vspace{-4mm}
\caption{An example of a code readability rule defined in SonarQube ASAT} 
\label{fig:example}
\end{figure}

The code readability rules defined in SonarQube are a subset of the issues raised in code smell rules. In Figure \ref{fig:example}, an example of a code readability rule defined in SonarQube ASAT is shown. The rule's description specifies that it is used to ensure code readability. Another example is the cognitive complexity metric \cite{Campbell2018}, which is related to some aspects of understandability \cite{Marvin2020}.

\section{Study Design}

\begin{figure*}[]
\centerline{\includegraphics[width=1.00\textwidth]{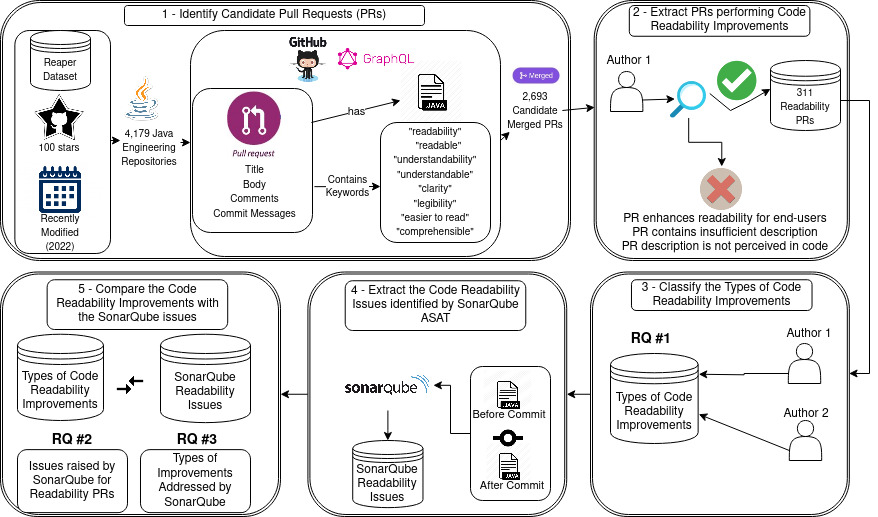}}
\vspace{-4mm}
\caption{Overall proposed architecture} 
\label{fig:architeture}
\end{figure*}

 The overall approach to answering the three research questions mentioned in previous section is illustrated in Figure \ref{fig:architeture}. The main steps involved in this approach are: (1) Identify the Candidate Pull Requests, (2) Extract PRs that perform Code Readability Improvements, (3) Classify the Types of Code Readability Improvements, (4) Extract the Code Readability Issues identified by SonarQube ASAT, and (5) Compare the Code Readability Improvements with the SonarQube issues. The details of each step are discussed in the following subsections. A replication package, comprising the dataset, scripts, classes, assessments, and instructions for reproduction is available \cite{carlos_eduardo_c_dantas_2022}.  

\subsection{Identify Candidate Pull Requests (PRs)}

To identify PR candidates related to code readability, we first selected non-forked Java repositories on GitHub using the dataset provided by Reaper \cite{Munaiah2017}. This tool searches for engineering repositories that exhibit evidence across seven dimensions, including collaboration among different Github users, effective issue management, and a history that demonstrates sustained evolution. These factors are crucial in our selection process for identifying readable PRs under a peer review system, preventing the inclusion of potentially trivial (i.e., toy) GitHub repositories in our dataset, which could result in inaccurate conclusions.  We limited our selection to Reaper-engineered repositories (score-based classifier) with at least 100 stars and one modification since the year 2022. This resulted in a set of 4,179 Java repositories.

Then, we mined merged PRs approved by the reviewer(s) for each of these 4,179 Java engineering repositories, using GitHub APIs such as the GitHub REST API\footnote{https://docs.github.com/en/rest} and GitHub GraphQL API\footnote{https://docs.github.com/en/graphql}. To identify merged PR candidates related to code readability, we searched for keywords such as ``readability``, ``readable``, ``understandability``, ``understandable``, ``clarity``, ``legibility``, ``easier to read``, and ``comprehensible`` in any of the PR fields including PR title, description, comments, and commit messages. These keywords were chosen based on their use in previous works \cite{Roy2020, Fakhoury2019}.

During the process, we discarded PRs that did not involve any changes to Java files in their commits. Some PRs made improvements to the readability of textual files such as \textit{readme} and \textit{.md} files. Additionally, PRs could have improved the readability of other programming language files since Java repositories may contain source code from languages like Kotlin, Scala, or Groovy. As a result, we were left with 2,693 merged PR candidates. Furthermore, we made sure that all remaining PRs had at least two distinct participants (the PR author and reviewer) to ensure a thorough peer code review process.

\subsection{Extract PRs performing Code Readability Improvements}

The first author of this paper manually analyzed each of the 2,693 merged PR candidates to identify the ones that resulted in improved code readability. The selection process involved evaluating each PR based on three objective criteria that needed to be met: 

\subsubsection{The PR enhances the code readability for other developers instead of end-users}  Some merged PRs were discarded because may modify Java files to improve the readability for end-users such as user interface (UI) by modifying console, log messages, UI element visualization or the inclusion of new features. For instance, consider PR \#264 from \textit{hibernate/hibernate-orm}, which proposes creating a new readable annotation called \textit{@IncrementGenerator}. While this new feature may improve readability for users of the Hibernate framework, it does not improve code readability for other developers seeking a better understanding of the internal source code of the Hibernate framework.

\subsubsection{The description provided by the developer explicitly reports the type(s) of code readability improvement that was performed} The detailed description is considered only if it is followed by one of the readability keywords used in this research. Some merged PRs were discarded because the developer or reviewer provided an insufficient description. For example,  \textit{PR \#874} from \textit{apache/avro} repository has a commit \textit{ef8f99e} with the description: \textit{``improve code readability``}. This description is insufficient to explain what type of readability improvement was performed, and therefore this PR was discarded.

\subsubsection{The code readability improvements should be evident in the source code's diffs within the commits} Although some merged PRs had a detailed description of the code readability improvement, they were discarded because these improvements were not discernible in any of the PR's commits. For instance, in the \textit{PR \#2659} from the \textit{azkaban/azkaban} repository, the reviewer recommended converting some literals into constants, but the developer did not implement these suggestions in the source code. Consequently, this PR was rejected.

\begin{figure}[]
\centerline{\includegraphics[width=0.50\textwidth]{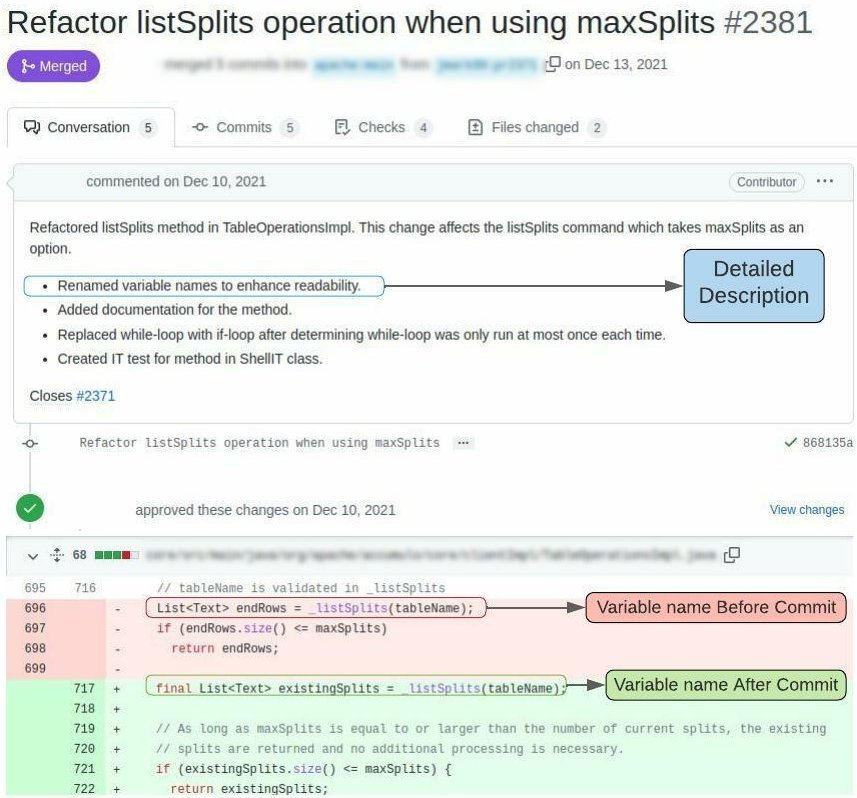}}
\vspace{-4mm}
\caption{PR \# 2381 from \textit{apache/accumulo} repository, with the code readability improvement and their respective code change (before and after the commit)} 
\label{fig:PRexample}
\end{figure}

Figure \ref{fig:PRexample} illustrates an example of a selected PR that improves code readability. In this PR, the developer improved code readability by renaming variable names, but he also made other code changes such as adding documentation for the method, creating tests, and replacing the while-loop with the if-loop. We verified the code readability improvement in commit \textit{868135a}, where the variable name \textit{endRows} was changed to \textit{existingSplits}. To collect data for this sample, we extracted the detailed description (\textit{``Renamed Variable names to Enhance Readability``}) and the corresponding code diff (before and after the commit) from the PR. Note that we only consider the changes explicitly mentioned by the developer or reviewer as code readability improvements and discard any modifications not related to readability. Our analysis is based solely on the developer's or reviewer's descriptions without any personal interpretation or addition of text, i.e., if the developer did not explicitly mention improvements related to code readability, we did not consider them.

In this step, we collected a total of 311 merged PRs that improved code readability, resulting in 400 candidate samples that included detailed descriptions and their corresponding code diffs. It is worth noting that a single PR could have multiple detailed descriptions regarding code readability improvements.  For instance, the \textit{PR \#2144} from \textit{RPTools/maptool} repository has the title: \textit{minor perf, readability improvements}. And then, the developer added 20 commits to this merged PR, with each commit message explicitly detailing the type of code readability improvement being made. For example, the commit \textit{33fe57c} had the message \textit{``remove redundant addAll()/putAll()``}, while another commit focused on other type of improvement: \textit{20d3935}, \textit{replacing manual min/max with Math.min()/max()}.

\subsection{Classify the Types of Code Readability Improvements Accepted by Reviewers}

To address RQ \textit{\#1}, two of the authors of this paper (each with over 15 years of experience in Java programming language) manually analyzed all 311 merged PRs, as previously described. Each sample includes the following details: (i) the hyperlink to the PR; (ii) the detailed descriptions of the code readability improvements provided by the developer or reviewer; (iii) the code diff for each improvement, with the before-commit source code in red font and the after-commit source code in green font; and (iv) the GitHub diff page link for each commit.

To code the collected samples, both authors independently analyzed the PRs and their corresponding candidate code readability improvements, assigning tags to each type of improvement implemented by the PR. To ensure consistency, the authors shared their tags in a common repository while analyzing the PRs. If an author found that a PR discussion was not related to code readability improvements, it was labeled as a \textit{``false positive``}.

After analyzing the 311 merged PRs and tagging the types of code readability improvements, the two authors discussed any conflicts in their tags and discarded any PRs where they couldn't come to an agreement. The final process resulted in 284 PRs (27 were discarded) with 370 types of code readability improvements across 109 Java repositories. The authors then grouped similar types of changes with similar purposes (e.g., add modifier \textit{final} and add annotation \textit{@Override}) to create the first taxonomy. The first draft of the taxonomy described the different purposes of code readability improvements, such as \textit{verbose}, \textit{clarify}, \textit{simplify}, \textit{formatting}, \textit{unused}, etc. The authors then refined the taxonomy by renaming some categories and moving sub-categories through the taxonomy. 

\begin{figure}[]
\centerline{\includegraphics[width=0.50\textwidth]{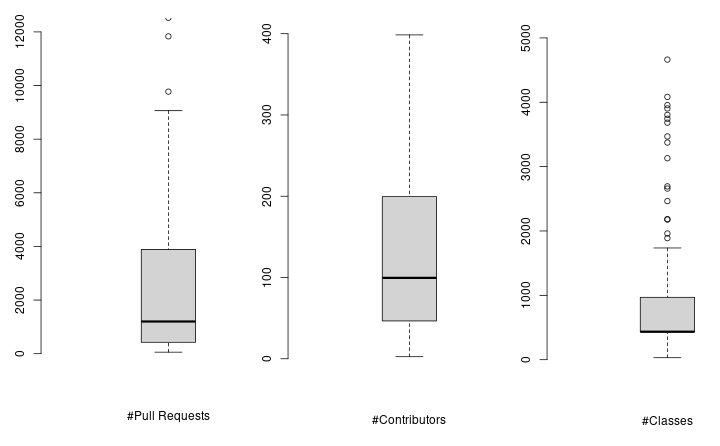}}
\vspace{-4mm}
\caption{Characteristics of the 109 Java repositories used in our study} 
\label{fig:distribution}
\end{figure}
\begin{figure}[]
\centerline{\includegraphics[width=0.42\textwidth]{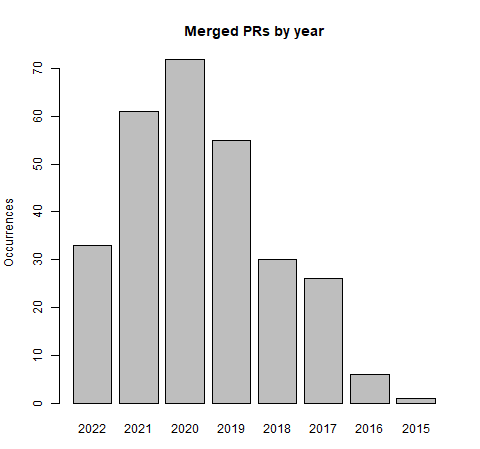}}
\vspace{-4mm}
\caption{Occurrences of merged PRs by year} 
\label{fig:pr_year}
\end{figure}

Figure \ref{fig:distribution} displays boxplots depicting the distribution of Pull Requests (PRs), contributors, and Java classes for the 284 merged PRs from 109 repositories. These PRs were contributed by 293 distinct developers, and 322 distinct reviewers analyzed them. Figure \ref{fig:pr_year} illustrates the distribution of all merged PRs grouped by year. The majority of the PRs were merged in recent years, indicating that our dataset reflects current trends and developers' views on code readability improvements.

\subsection{Extract the Code Readability Issues identified by SonarQube ASAT}

In this step, for each of the 370 code readability improvements used in RQ \#1, we collected two versions (before and after the commit) of the Java file that contains the improvements described by developers. We then created a client script to consume the SonarQube core interface, which enabled us to retrieve the issues (i.e., rule violations) associated with each class.  We selected SonarQube as our primary ASAT tool due to its widespread adoption in both industrial settings (with over 80,000 projects) and academic research (being employed in numerous prior works). Additionally, SonarQube incorporates rules from other ASATs, such as FindBugs and PMD, making it a comprehensive choice for Java projects.

Through this process, SonarQube detected a total of 161 distinct issues. To identify only those related to rule violations of code readability, we analyzed the description of each issue using the same readability keywords described in previous section that were used to filter merged PRs. After this filtering process, we found that 51 out of the 161 issues were related to code readability, while the remaining 110 issues were not related and were therefore discarded. 

This step results in the identification of two sets of SonarQube issues. The first set refers to issues that were detected in the Java class before the developer performed the code readability improvements that he described in the PR (i.e., before the commit), while the second set pertains to issues identified after the readability improvement (i.e., after the commit).

\subsection{Compare the Code Readability Improvements with the SonarQube Issues}

To address RQ \#2, we developed a script to systematically count Java classes and instances associated with each SonarQube rule that raised an issue, for each set (before and after the commit). In this context, \textit{``instances``} refer to the number of occurrences of each specific issue. It's worth noting that a class might have multiple occurrences of the same issue.  To address RQ \#3, we extracted all the affected code snippets from Java classes inside the before commit set, for each SonarQube rule related to code readability that were raised. These snippets were then compared with the code diffs performed by developers from the 370 code readability improvements used in RQ \#1. Our focus was on selecting only cases where SonarQube recommended the same code readability improvement that was described and implemented by the developers. 

\section{Results}

In this section, the results will be shown according to each research question.

\subsection{\textbf{RQ \#1) What types of code readability improvements do developers describe and perform in Pull Requests (PRs)?}}

Table \ref{tab:rq1} reports the types of code readability improvements found in the analyzed PRs, grouped into seven categories. We only list types of code readability improvements observed at least five times. The remaining observed improvements are classified as ``others``.

\subsubsection{Clarify Code Intent} (93 out of 370 samples) - this category aims to improve the code to be more intuitive and self-describing, reducing the subjectivity of other developers in reading and understanding code. In this category, the code readability improvements do not necessarily remove additional tokens as the \textit{Reduce Code Verbosity} category.

\textit{Improve Naming} (43 out of 93 samples) - this type of improvement was already perceived in previous works \cite{Fakhoury2019} \cite{Pantiuchina2020}, and shows how the quality of identifiers names are important to developers, aiming to express their own intents inside the project. We noticed 20 samples renaming variables, 18 renaming methods, and 5 renaming others (class, attribute, or constant). Examples:
\begin{itemize}
    \item \textit{PR \#2381} from \textit{apache/accumulo} repository: \textit{``Renamed variable names to enhance readability.``}
  \item \textit{PR \#647} from \textit{gdg-x/frisbee} repository: \textit{``Renaming base method to improve readability.``}
\end{itemize}   

\textit{Replace Magic Literals} (18 out of 93 samples) occurs when the literals values whose meaning might not be obvious to the reader of the code \cite{Anquetil2022}. The \textit{PR \#4337} from \textit{stripe/stripe-android} repository replaces a method parameter with value \textit{``429``} with the constant \textit{``HTTP\_TOO\_MANY\_REQUESTS``}. The developer explained his change: \textit{``Replace plain Int/String values for constants for readability``}. 

\textit{Include Modifiers and Annotations} (13 out of 93 samples)  aim to expose the scope of the elements in a class. For example, when a developer includes the \textit{``final``} modifier in a variable, other developers understand the variable is immutable. The same idea is applied for \textit{@NotNull} annotation on non-null fields or \textit{@Override} annotation for overwriting methods from the superclass. 

\textit{Replace Generic with Specific APIs} (12 out of 93 samples) described \textit{``specific``} APIs to explicitly express the code intent. These samples has descriptions such as  \textit{``provide more clarity``}, \textit{``refactor to specialized functional interface``}, \textit{``improve code clarity using immutable interfaces instead of the generic collection interface``}, and \textit{``improve the readability of tests, and also be more specific``}. In these scenarios, we observed a consistent pattern where developers opted to replace a versatile API capable of serving multiple purposes with a specific API that effectively conveys intent.
 
 For example, The \textit{PR \#2144} from \textit{RPTools/maptool} repository replaces a \textit{Arrays.asList()} to \textit{Collections.singletonList()} because the variable was using only one element. In other words, employing \textit{Collections.singletonList()} aids comprehension by explicitly indicating that the variable will solely contain a single element, i.e., \textit{Arrays.asList()} is the generic API, and the \textit{Collections.singletonList()} is the specific API. Another example is the \textit{PR \#134} from \textit{jenkinsci/plugin-installation-manager-tool} repository, where the developer refactored the unit tests to consume the AssertJ library, which uses the fluent interface method \cite{Nakamaru2020}. 







\begin{table*}[htbp]
\centering
\caption{Code readability improvements performed by Developers}
\vspace{-3mm}
\label{tab:rq1}
\begin{tabular}{l|l|cl|cl}
\hline
Category & Type of Code Readability Improvement & \multicolumn{2}{l}{Occurrences} & \multicolumn{2}{l}{Total} \\
\hline
\multirow{ 5}{*}{Clarify Code Intent} 
               & Improve Naming & 43 & \Chart{0.46}               
               & \multirow{ 5}{*}{93} & \multirow{ 5}{*}{\Chart{0.25}} \\     
               & Replace Magic Literal & 18 & \Chart{0.19} \\
               & Include Modifiers and Annotations & 13 & \Chart{0.14}  \\     
               & Replace Generic with Specific APIs & 12 & \Chart{0.13}  \\     
               & Others & 7 & \Chart{0.08} \\
\hline
\multirow{ 7}{*}{Reduce Code Verbosity} 
               & Remove Redundant (Boilerplate) Operations & 26 & \Chart{0.3}             
               & \multirow{ 7}{*}{86} & \multirow{ 7}{*}{\Chart{0.23}}  \\ 
               & Replace Custom Implementation to Consume Existing Java API & 24 & \Chart{0.28} & & \\
               & Replace Anonymous Class with Lambda  &  12 & \Chart{0.14} & & \\               
               & Simplify Loops with Lambda/for-each  &  10 & \Chart{0.12} & & \\
               & Simplify Redundant try/catch blocks  &  6 & \Chart{0.07} & & \\               
               & Prefer Static Imports & 5 & \Chart{0.06} & & \\                     
               & Others & 3 & \Chart{0.03} \\ 
\hline
\multirow{ 3}{*}{Improve Code Modularity} 
               & Extract Methods &  41  & \Chart{0.71}              
               & \multirow{3}{*}{57} & \multirow{3}{*}{\Chart{0.15}}  \\ 
               & Extract Classes & 11 & \Chart{0.2} & & \\ 
               & Others & 5 & \Chart{0.08}   \\
\hline
\multirow{ 6}{*}{Simplify if/else conditions} 
               & Convert Negative Conditions into Positive Ones  &  10  & \Chart{0.22}
               & \multirow{ 6}{*}{45}  & \multirow{ 5}{*}{\Chart{0.12}} \\
               & Split Single Line If Statement into Multiline If Statements & 10 & \Chart{0.22} \\
               & Reduce Nested if-else Depth &  8  & \Chart{0.18} \\
               & Replace If/else chains with Switch & 8  & \Chart{0.18} \\         
               & Prefer the If/else Code Style Pattern & 5 & \Chart{0.11} \\
               & Others & 4 & \Chart{0.08}  \\    
\hline
\multirow{ 4}{*}{Remove Unused Code} 
               & Remove Unused Conditions/Loops & 9 & \Chart{0.32} \\
               & Remove Unused Method Parameters  &  7 & \Chart{0.25}
               & \multirow{ 4}{*}{28} & \multirow{ 4}{*}{\Chart{0.08}} \\       
               & Remove Unused Methods/Fields & 6 & \Chart{0.21} \\              
               & Remove Unused Test Asserts and Exceptions & 6 & \Chart{0.21} \\                          
\hline
\multirow{ 3}{*}{Improve Formatting (Source code structure)} 
               & Break long lines  &  11 & \Chart{0.41}
               & \multirow{ 3}{*}{27} & \multirow{ 3}{*}{\Chart{0.07}} \\
               & Move Method/Field inside class & 8 & \Chart{0.30} \\
               & Improve Code Indentation & 8 & \Chart{0.30} \\
\hline
\multirow{ 2}{*}{Improve Comments} 
               & Include new Comments (and Javadoc)  &  13 & \Chart{0.57}
               & \multirow{ 2}{*}{23} & \multirow{ 2}{*}{\Chart{0.06}} \\
               & Clarify comments (and Javadoc) & 10 & \Chart{0.43} \\
\hline               
Others & & & & 11 & \Chart{0.02} \\
\hline               
\multicolumn{4}{c}{\textbf{Total}} & \textbf{370} & \Chart{1.00} \\
               
\hline

\end{tabular}
\vspace{-4mm}
\end{table*}

\subsubsection{Reduce Code Verbosity} (86 out of 370 samples). In this category, the developers described code verbosity as an unconcise code, i.e., using unnecessary additional tokens to solve a task. The reduction of code verbosity eliminates unnecessary tokens, writing more concise code. For example, in  \textit{PR \#965} from \textit{broadinstitute/picard} repository, the developer wrote in the commit \textit{\#bdb0eab}: \textit{replace a verbose comparator definition with a shorter one line}. This commit replaced the \textit{RepresentativeReadComparator} class with 7 LOC (lines of code) to a new single-line comparator using the lambda feature.

\textit{Remove Redundant (Boilerplate) Operations} (26 out of 86 samples) occurs when the developer writes extra unnecessary code to implement a feature. For example, the \textit{PR \#2144} from \textit{RPTools/maptool} repository removes redundant type casting, as \textit{Shape s = (Shape) a;} replaced to \textit{Shape s = a;}. In this case, the casting to (Shape) is an unnecessary extra effort, using additional tokens to read and understand the same task.

\textit{Replace Custom Implementation to Consume Existing Java API} (24 out of 86 samples) occurs when the developer writes an algorithm to implement some feature that was already defined on some Java API. The \textit{PR \#45936} from \textit{elasticsearch} repository replaced a 6 LOC custom encoding/decoding vector to consume a single-line \textit{java.nio.ByteBuffer.wrap()} method.

The lambda expressions are widely used in two types of code readability improvements:  \textit{Replace Anonymous Class with Lambda} (12 out of 86 samples) and  \textit{Simplify Loops with Lambda/for-each} (10 out of 86 samples). This is because developers can avoid the verbosity by reducing LOC. A comment on \textit{PR \#4703} in the apache/pulsar repository provides an example of how lambda expressions can reduce the effort required to write code. The comment states, \textit{``The anonymous classes were converted to lambda in places where it improves readability and reduces the lines of code.``}

The \textit{Simplify Redundant try/catch blocks} (6 out of 85 samples) occurs generally to reduce LOC by using \textit{try-with-resource} statement, i.e., a try statement that declares one or more resources, or using multi-catch block, i.e., handling multiple exceptions in a single catch block, avoiding redundant code to handle similar exceptions. The \textit{Prefer Static Imports} (5 out of 85 samples) uses static imports to access the static members of a class directly without writing the class name. The \textit{PR \#4683} from \textit{apache/pulsar} illustrates by the following comment: \textit{``Add static import statements for Assert to improve the readability of the tests``}. This change replaces \textit{Assert.assertEquals()} to \textit{assertEquals()} in many lines of code.

\subsubsection{Improve Code Modularity} (57 out of 370 samples) - modularity separates the program complexity into smaller parts \cite{Beck2011}. Most of the detailed descriptions given by developers mentioned the Extract Method (41 of 57 samples) and Extract Class (11 of 57 samples) refactoring operations to reduce unreadable long methods. Examples:
\begin{itemize}
    \item \textit{PR \#710} from \textit{undertow} repository: \textit{``Second commit refactors URLDecodingHandler into smaller methods for readability.``}
    \item \textit{PR \#378} from \textit{jenkinsci/remoting} repository: \textit{``The first PR extracts a few methods for improved readability. This makes it easier to isolate the different operations and see what is going on.``} 
    \item \textit{PR \#383} from \textit{redhat-developer/intellij-tekton} repository: \textit{``extracted smaller methods to improved readability/maintainability``}  
\end{itemize}      

We noticed 9 out of the 41 extract methods were also performed to remove duplicate code, and 5 out of 41 to reduce the \textit{``long parameter list``} code smell. The Extract Class refactoring was applied when the extracted code was outside the scope of the class. The \textit{PR \#1901} from \textit{CorfuDB} repository illustrates by the following comment: \textit{``Separates log unit cache into a new class for more readability and testability.``}

\subsubsection{Simplify if/else conditions} (45 out of 370 samples) - we found a considerable amount of code readability improvements focused on clean up  if/else conditions. The \textit{Convert Negative Conditions into Positive Ones} (10 out of 45 samples) improvement refactored the if statements to remove the \textit{not operator (!)} in conditions. For example, the \textit{PR \#1303} from \textit{broadinstitute/picard} repository illustrates this removal in the comment: \textit{``I think it's clearer to have flags that are in a positive state, not a negative state.``}

The \textit{Split Single Line If Statement into Multiline If Statements} (10 out of 45 samples) divides the conditions separated by operators \textit{(``$\&\&$``)} or \textit{($``||``)$} into a \textit{if-else} chain. The \textit{Reduce Nested If-else Depth} (8 out of 45 samples) aims to reduce the nuanced sequence of conditionals. For example, the \textit{PR \#1508} from \textit{forcedotcom/SalesforceMobileSDK-Android} repository reduce the nested if-else nested depth from 3 (before a commit) to 1 (after a commit), followed by the comment: \textit{``More readable validation conditionals.``}

In this category, we also reported \textit{``Replace If/else chains with Switch``} (8 out of 45 samples) and \textit{Prefer the If/else Code Style Pattern} (5 out of 45 samples), i.e., conditions where the developer did not follow the convention. In \textit{PR \#1753} from \textit{CorfuDB} repository, the reviewer asked for changes: \textit{``I prefer not to use one-liner if condition. Also, PMD believes it's a bad practice``}.

\subsubsection{Remove Unused Code} (28 out of 370 samples) - while the \textit{Reduce Code Verbosity} category removes unnecessary additional code to implement a feature, this category removes unused code, e.g., a custom method never called, or a field never used. For example, the \textit{PR \#2206} from \textit{apache/hive} repository illustrates with the comment: \textit{``Remove unused fields/methods and deprecated calls in HiveProject. Why are the changes needed? Improve readability.``}

The \textit{Remove Unused Conditions/Loops} (9  out of 28 samples) removes \textit{ifs/else} conditions or \textit{loops} never used. The \textit{PR \#1604} from \textit{apache/accumulo} illustrates with the comment: \textit{``ThriftServerBindsBeforeZooKepperLockIT contains three 'while' blocks that never loop. The outer loop does not appear to be necessary in each of these cases. These loops have been removed to increase the readability of the code.``} 

\subsubsection{Improve Formatting (Source Code Structure)} (27  out of 370 samples). Previous readability models have already addressed code formatting as line length and indentation \cite{Buse2010} \cite{Posnett2011} \cite{Dorn2012AGS}. This improvement aims to visually group related code blocks and make them easier to distinguish from one another.

\textit{Break long lines} (11  out of 27 samples) limits the number of characters per line. The \textit{PR \#6013} from \textit{jenkins} repository illustrates with the comment: \textit{``Applies the rectangle rule to some extremely long lines to improve readability and sets the maximum line length at 240 columns.``}

\textit{Move Method/Field inside Class} (8 of 27 samples) tries to organize the order of the elements (attributes, methods) inside the Java class. The  \textit{PR \#67175} from \textit{elasticsearch} repository illustrates with the comment: \textit{``Improve readability of Node transform/forEach typed methods by moving the type token to the front of the method, before the lambda declaration.``}

\subsubsection{Improve Comments} (23 of 370 samples). Aims to improve the readability of code via comments \cite{Stapleton2020}. We find 13 code readability improvements creating new comments (most of them using the Javadoc format). And the remaining 10 samples asked for improving the text quality. Examples:

\begin{itemize}
    \item \textit{PR \#931} from \textit{jenkinsci/git-plugin} repository: \textit{``Add Javadoc comments for better readability.``}
     \item \textit{PR \#20353} from \textit{gradle} repository: \textit{``Rewrites comment to read better Readability is increased which will allow for faster comprehension of the edited file.``}
\end{itemize}  


Our findings indicate that some types of code readability improvements can be recommended by readability models and/or tools. For instance, (\textit{``Break Long Lines``, ``Improve Code Indentation``, ``Include new Comments``, ``Improve Naming``, ``Prefer the If/else Code Style Pattern``}) have already been used as metrics in code readability models \cite{Buse2010} \cite{Dorn2012AGS}. Other models could identify a complex and unreadable line of code, e.g., \textit{``Split Single Line If Statement into Multiline If Statements``} \cite{Scalabrino2018}. The cognitive complexity metric use \cite{Campbell2018} can suggest minimizing the complexity of control flows (e.g., \textit{``Reduce Nested If/else Depth``, ``Simplify Redundant Try/catch Blocks``, ``Simplify Loops using Lambda/for-each``}). Some tools can detect bad smells such as \textit{Large Class} and \textit{Large Method}, suggesting the refactoring to extract a class or a method.

Nonetheless, our findings indicate that some opportunities for improvement could be  challenging to be detected by  code readability tools. For example, the \textit{``Replace Generic with Specific APIs``} could be an individual developer's decision based on his/her own experience. In some of our observations,  \textit{java.util.Map} was replaced with \textit{com.google.common.collect.ImmutableMap}, or \textit{java.util.ArrayDeque} with \textit{java.util.concurrent. ConcurrentLinkedDeque}. Another example is the \textit{Include Modifiers and Annotations}, where an annotation or a modifier could both be considered a boilerplate code if the variable does not need them. For example, the \textit{PR \#3152} from \textit{apache/pinot} repository asked to remove the unused \textit{@Nonnull} annotation to keep the readability.

\noindent
\begin{center}
\fbox{\begin{minipage}{25em}
\textbf{RQ \#1 Answer:} Through our analysis, we have created a comprehensive catalog of 26 different code readability improvements that fall under seven distinct categories. Our findings indicate that in most of the cases, developers were able to improve code readability by making it more intuitive, modular, and concise. While some of the changes made were subjective in nature, they pose a significant challenge for new readability tools and models to replicate.

\end{minipage}}
\end{center}
\vspace{.2em}

\begin{table*}[htbp]
\centering
\caption{Top 25 SonarQube Issues Detected Before and After the Code Readability Improvement Performed by Developers}
\vspace{-3mm}
\label{tab:rq2}
\begin{tabular}{l|l|c|c|c|c}
\hline & & \multicolumn{2}{c|}{Issues Before Commit} & \multicolumn{2}{c|}{Issues After Commit}\\
\cline{3-6} \raisebox{1.2ex}{Rule} & \raisebox{1.2ex}{SonarQube Description} & Classes & Instances & Classes & Instances \\
\hline
java:S3776 & Cognitive Complexity of methods should not be too high & 132 & 396 & 124 & 368 \\
java:S1192 & String literals should not be duplicated & 108 & 635 & 104 & 615 \\
java:S125 & Sections of code should not be commented out & 56 & 225 & 54 &  217 \\
java:S1117 & Local variables should not shadow class fields  & 52 & 132 & 53 &  132 \\
java:S2293 & The diamond operator $("<>")$ should be used & 51 & 274 & 46 &  265 \\
java:S1066 & Collapsible ``if`` statements should be merged  & 44 & 88 & 42 &  87 \\
java:S135 & Loops should not contain more than a single ``break`` or ``continue`` statement & 39 & 86 & 39 &  84 \\
java:S1604 & Anonymous inner classes containing only one method should become lambdas & 36 & 129 & 31 &  104 \\
java:S1141 & Try-catch blocks should not be nested & 35 & 87 & 33 &  85 \\
java:S106 & Standard outputs should not be used directly to log anything & 26 & 98 & 24 &  98 \\
java:S1124 & Modifiers should be declared in the correct order & 24 & 115 & 24 &  116 \\
java:S1125 & Boolean literals should not be redundant & 23 & 170 & 22 &  167 \\
java:S1659 & Multiple variables should not be declared on the same line & 21 & 47 & 20 &  44 \\
java:S1155 & Collection.isEmpty() should be used to test for emptiness  & 19 & 35 & 18 &  34 \\
java:S117 & Local variable and method names should comply with a naming convention & 18  & 53 & 17 &  50 \\
java:S116 & Field names should comply with a naming convention & 17 & 237 & 18 &  239 \\
java:S1612 & Lambdas should be replaced with method references & 17 & 33 & 17 &  26 \\
java:S1450 & Private fields only used as local variables in methods should become local variables & 16 & 21 & 17 &  24 \\ 
java:S3008 & Static non-final field names should comply with a naming convention & 16 & 31 & 15 & 28 \\
java:S1602 & Lambdas containing only one statement should not nest this statement in a block & 15 & 35 & 17 & 38 \\
java:S1126 & Return of boolean expressions should not be wrapped into an ``if-then-else`` statement & 15 & 25 & 13 & 19  \\
java:S1611 & Parentheses should be removed from a lambda input parameter when its type is inferred & 14 & 26 & 15 & 27 \\
java:S1144  & Unused ``private`` methods should be removed & 13 & 18 & 14 & 17 \\
java:S1121 & Assignments should not be made from within sub-expressions & 11 & 11 & 11& 11 \\
java:S6213 & Restricted Identifiers should not be used as Identifiers & 11 & 39 & 10 & 37 \\
\hline


\end{tabular}
\vspace{-4mm}
\end{table*}

\subsection{\textbf{RQ \#2) Do developers fix code readability issues identified by SonarQube in readability PRs?}}

Table \ref{tab:rq2} reports the top 25 SonarQube issues identified both before and after developers performed code readability improvements. The classes column indicates the number of classes in which the issue was raised by SonarQube, while the instances column indicates the total number of times the issue was raised by SonarQube.

Based on our findings, it appears that developers often do not address code readability issues identified by SonarQube. Specifically, we found that rule \textit{java:S3776}, which pertains to methods with excessive complexity that should be extracted, was violated in 396 methods from 132 classes. Even after code readability improvements were made, however, 368 methods from 124 classes still exhibited high complexity and failed to comply with the rule. Of the 25 rules analyzed, 10 showed an increase or no change in the number of classes with issues, while the remaining 15 rules exhibited a slight decrease in the number of affected classes.

While this outcome was somewhat anticipated given that developers typically only make changes to the code that was described in the PR, it is worth noting that there are many other opportunities to enhance code readability that often go unaddressed, suggesting that developers tend to focus on a limited set of specific improvements when tackling code readability issues. For instance, in PR \textit{\#793} from the \textit{apache/accumulo}, the developer removed unused methods and replaced loops with lambda expressions in the \textit{TabletServer.java} class. However, SonarQube identified issues across nine different rules, such as the use of magic literals, nested try-catch blocks and others. The class contained 3,630 lines of code prior to the commit, and 3,615 lines after the commit.

\noindent
\begin{center}
\fbox{\begin{minipage}{25em}
\textbf{RQ \#2 Answer:} Our analysis revealed that, in many cases, developers do not address  readability issues flagged by SonarQube. Although we cannot conclude if they neglect them, either because they   rate them as unimportant, or because they are unaware of them, we found that developers tend to focus on improving  readability in certain areas, while neglecting other aspects of the code that also would require attention.

\end{minipage}}
\end{center}
\vspace{.3em}

\subsection{\textbf{RQ \#3) Can SonarQube detect the code readability improvements performed by developers in PRs?}} 

To our surprise, we observed that SonarQube recommended the code readability improvements performed by developers only in 26 out of 370 code readability improvements (7,02\%). Some improvements were not expected to receive recommendations from SonarQube due to the subjective nature of the task, such as improving naming for method (``java:S100``), variable (``java:S117``), parameter (``java:S119``), static field (``java:S3008``), and constant naming (``java:S115``), which SonarQube only recommends when the names do not conform to conventions (such as the \textit{camelCase} pattern in Java), and most of the 43 instances of naming improvements we found in RQ \#1 replaced unclear names with more descriptive ones. 

In some cases, we observed discrepancies between the recommendations made by SonarQube rules and the way developers perceive code readability issues. For instance, SonarQube's ``java:S1192`` rule advises the removal of duplicated String literals by creating a constant to reference the String. However, in RQ \#1, we noticed that developers created constants to remove magic literals regardless of whether they were duplicated in the code. Another example is the java:S3776 rule for cognitive complexity, which SonarQube only flags if a method's control flow statements reach a certain level of complexity. However, in RQ \#1, we observed instances where developers had extracted large methods or classes without control flows, which SonarQube did not flag as problematic.

However, in some cases, SonarQube is unable to detect all instances of a particular issue. For instance, the ``java:S1604`` rule, which recommends converting anonymous inner classes to lambdas, is semantically equivalent to our ``Replace Anonymous Class with Lambda`` change type. Despite this, SonarQube only flagged the issue in three out of nine occurrences. 

Finally, we did not observe any SonarQube rules related to formatting, such as line length limits or simplifying loops using lambda expressions or for-each statements. This result highlights the potential for improvement in ASAT code readability rules as they are perceived by developers.

\noindent
\begin{center}
\fbox{\begin{minipage}{25em}
\textbf{RQ \#3 Answer:} Only 26 out of the 370 code readability improvements implemented by developers were flagged as issues by SonarQube. Our observations revealed instances where SonarQube had the relevant rule but did not raise the issue, as well as types of improvements that SonarQube does not have rules for, indicating potential areas for ASAT improvements.

\end{minipage}}
\end{center}
\vspace{.2em}

\section{Discussion}

This section discusses the implications driven from our study for researchers and developers, and the threats to validity.

\subsection{Implications for Developers and Researchers}

For researches, we have identified that SonarQube does not capture certain readability improvements made by developers. This presents an opportunity for improvement in ASATs, such as recommending more meaningful names instead of solely raising issues based on names that do not adhere to regular expressions.

Furthermore, the catalog comprising 26 distinct code readability improvements performed by developers during code review processes could provide valuable insights for enhancing readability models and ASATs. Despites some of these improvements were already observed in previous works (e.g., improve naming), others can offer a new set of features that align better with developers' perceptions of what constitutes readable code. 

Finally, in our sample, we observed that the majority of developers implemented a limited number of specific readability improvements within their PRs. To clarify, it is uncommon to come across PRs where developers thoroughly review the readability within the entire class. This finding highlights an opportunity to enhance the code review process by suggesting readability improvements that developers may have overlooked.

For developers, specifically Java programmers, our samples demonstrate that developers frequently prioritize enhancing code readability to make it more intuitive, modular, and concise. We have observed a notable emphasis on produce a more intuitive code by improving variable and method names, simplifying control flows, reducing verbosity in the code, and improve the modularity.

Our work also identified several cases where the developers improve code readability by using new features under the Java programming language evolution. Examples: \textit{try-with-resources}, \textit{lambdas}, \textit{pattern matching} and others. These findings emphasize the significance of embracing and familiarizing oneself with these features to enhance code readability and take advantage of the language advancements.

Finally, although our research identified numerous instances where ASATs did not recommend the code readability improvements made by developers, the tool could still offer valuable insights by suggesting code readability improvements that developers might not have perceived on their own.

\subsection{Threats to Validity}

\textit{Dataset creation}: In order to create our dataset, both the first and second authors manually reviewed the PRs and classified the types of code readability improvements. However, only the first author was responsible for the process of discarding merged PR candidates without code readability improvements. While this process may be more prone to discarding false negatives (i.e. potentially discarding PRs that contain code readability improvements), the two authors worked together to mitigate possible false positives (i.e. selecting PRs without code readability improvements). Unlike previous works that used randomly-stratified samples of PRs \cite{Pantiuchina2020}, we investigated all merged PR candidates to ensure the completeness of our dataset.

\textit{Developers Reliability}: We did not conduct a study to gather information on the developers' experience in improving code readability in this work. However, we mitigated this limitation by selecting only those PRs that involved at least two participants (an author and one or more reviewers), which underwent a peer-review process to improve code quality \cite{Ackerman1989}. Furthermore, our study analyzed discussions from over 600 professionals (293 authors and 322 reviewers), making it a significant sample size for a qualitative study.

\textit{Generalizability of the Findings}: Our study focuses exclusively on Java GitHub-hosted open-source projects. As such, the generalizability of our findings to other programming languages and platforms may be limited.

\textit{Selection of Candidate Repositories}: GitHub hosts a vast number of repositories, and a different dataset may produce varying results. For our study, we chose Java engineering Reaper repositories that had recently been modified and had 100+ stars. It is worth noting that previous research found that such repositories have 82\% precision and 83\% recall, while stargazers-based classifiers exhibit 96\% precision and 27\% recall \cite{Munaiah2017}. Despite this limitation, our dataset includes a substantial number of pull requests, contributors, and classes, as illustrated in Figure \ref{fig:distribution}.

\section{Related Work}

Previous studies have explored developer perceptions of code quality issues. For instance, Palomba and colleagues conducted a study to investigate developers' perceptions of code smells \cite{Fowler1999}, which refers to the extent to which developers consider code smells as serious design issues and the possible gaps between theory (i.e., what is considered a problem) and practice (i.e., what is actually a problem for developers). The experiment involved showing classes with code smells to developers and asking them if the code exhibited any design or implementation problems. The researchers analyzed the developers' responses to understand their interpretations of potential design issues in the code  \cite{Palomba2014}.

The developer's perceptions and definitions are associated with their personal opinions based on a subject (code smells,  architectural roles, or class co-changes) about a source code developed by them. Some of these personal opinions could have differences from the theory about the subject. This research has a  similar purpose, i.e., extracting perceptions from developers about code readability improvements based on their own texts on PRs about their changes. If one developer prefers to add comments and another developer prefers to remove them, there are two different perceptions about the effects of the comments on code readability.

Some recent work tried to evaluate if state-of-art readability models are able to capture readability improvements as explicitly mentioned by developers in their PR descriptions or commit messages. Pantiuchina et al. investigated code quality metrics that measure cohesion, coupling, complexity, and readability in 1,282 valid commits \cite{Pantiuchina2018} comparing the metric values before and after each commit. Their analysis concludes that the metrics could not capture the developer's interpretation of code quality, and the hints from metrics should be complemented by developers' feedback. Our research searches for developer feedback inside PRs to find their changes in improving code readability. 

The most related research to our work is from Fakhoury et al. \cite{Fakhoury2019}, which investigates readability improvements on 548 commits and 2,323 classes manually looking at the commit messages and source code. Their work confirms other previous work, i.e., the readability metrics could not capture the readability improvements on commits. Moreover, they find that complexity metrics as \textit{McCabe} \cite{McCabe1976} decrease on readability improvements, and refactoring operations as extract method are often on readability commits. Our work has two main differences: first, we search for qualitative types of code readability improvements instead of metric values. For this, we consider PRs instead of only committing messages to obtain more information from developers under a revised PR. The second difference involves the readability improvements using statistical tests. Instead of using all classes under a unique test, we split the classes under the types of code readability improvements performed by developers.

Previous works proposed studies involving developers' motivations to perform refactoring operations, some of them related to readability. Pantiuchina and colleagues manually analyzed 551 merged PRs and they find 468 from 1,117 instances with readability improvements \cite{Pantiuchina2020}, which evidences a correlation between readability improvements and refactoring operations. The readability improvements are related to renaming variables and cleanup code. Our work uses a similar qualitative analysis methodology, but we directly search for PRs where developers described code readability improvements. Other works \cite{Silva2016} used Thematic analysis \cite{Cruzes2011} to extract and produce higher-order themes \cite{Bradley2007QualitativeDA}.

Several works have mined code from merged PRs. Coelho et al. mined refactoring-inducing PRs, i.e., refactoring operations performed in commits after the initial PR commit until the merge commit is added to the repository. The commits performed in this interval result from the interaction between the developers and reviewer discussions \cite{Flavia2021}. They found that 30.2\% of refactoring-inducing operations in 1,845 PRs, and refactoring-inducing has some characteristics, like more time to merge, a number of reviewers, more discussion (PR comments), and more file changes. Our work uses merged PRs but aimed to collect types of code readability improvements instead of refactoring-inducing operations. 

\section{Conclusions}

In this paper, we conducted a qualitative analysis to explore how developers perform code readability improvements, with the goal of comparing recommendations provided by ASATs, such as SonarQube, and the  improvements developers actually have performed. So, we  extracted the source code changes to improve readability implemented in PRs by developers, and used them to investigate whether developers' perceptions of code readability are already addressed by an existing ASAT and if the ASAT recommendations have already been identified and performed by developers.

The main contribution is a catalog of 26 actual types of code readability improvements implemented by developers and merged into the codebase. We also found a prevalence in refactoring code to be more intuitive, modular, and less verbose. Interestingly, some of  types of improvements, such as formatting or refactoring loops to for-each and lambda, are not yet addressed in ASAT rules, indicating a valuable opportunity for improving ASAT rules in line with developers' perceptions.

This study presents several opportunities for future research. Firstly, it could be extended to investigate code readability improvements in other programming languages. Less verbose languages might have different types of code readability improvements that are more prevalent. Secondly, a more detailed investigation could be conducted to understand the reasons why ASAT cannot raise issues for some classes that contain rule violations. Such insights could help improve the accuracy of such tools. This study could also be expanded by incorporating other ASAT like Checkstyle, SpotBugs, and others.


\bibliographystyle{IEEEtran}
\bibliography{ieee}

\vspace{12pt}
\color{red}

\end{document}